\documentclass[reprint,footnoteinbib,prl,twocolumns]{revtex4-1}
\usepackage{amsfonts,amsmath,amssymb}
\usepackage[pdftex]{graphicx}
\usepackage{mathrsfs}
\usepackage{wasysym}
\usepackage{bm}

\usepackage[svgnames]{xcolor}
\usepackage[
	colorlinks=True,linkcolor=DarkRed,citecolor=ForestGreen,urlcolor=MediumBlue,
	pdfstartview=FitH,bookmarks=False,pdfpagemode=UseNone
]{hyperref}

\begin{document}

\title{Peltier Cooling of Fermionic Quantum Gases}

\author{Ch.~Grenier} 
\affiliation{Institute for Quantum Electronics, ETH Z\"urich, 8093 Z\"urich, Switzerland}

\author{A.~Georges}
\affiliation{Coll\`ege de France, 11 place Marcelin Berthelot, 75005 Paris, France.}
\affiliation{Centre de Physique Th\'eorique, Ecole Polytechnique, CNRS, 91128 Palaiseau Cedex, France.}
\affiliation{DPMC, Universit\'e de Gen\`eve, CH-1211 Geneva, Switzerland.}

\author{C.~Kollath}
\affiliation{HISKP, University of Bonn, Nussallee 14-16, D-53115 Bonn, Germany}

\date{\today}

\begin{abstract}{
We propose a cooling scheme for fermionic quantum gases, based on the principles of the Peltier thermoelectric effect and energy filtering.  
The system to be cooled is connected to another harmonically trapped gas acting as a reservoir. 
The cooling is achieved by two simultaneous processes:
(i) the system is evaporatively cooled and (ii) cold fermions from deep below the Fermi surface of the reservoir 
are injected below the Fermi level of the system, in order to fill the 'holes' in the energy distribution. 
This is achieved by a suitable energy dependence of the transmission coefficient connecting the system to the reservoir. 
The two processes can be viewed as simultaneous evaporative cooling of particles and holes. 
We show that both a significantly lower entropy per particle and faster cooling rate can be achieved 
than by using only evaporative cooling. 
%
}
\end{abstract}
\maketitle

\paragraph{Introduction}

\begin{figure}[ht!]
  \includegraphics[width=1.0\linewidth]{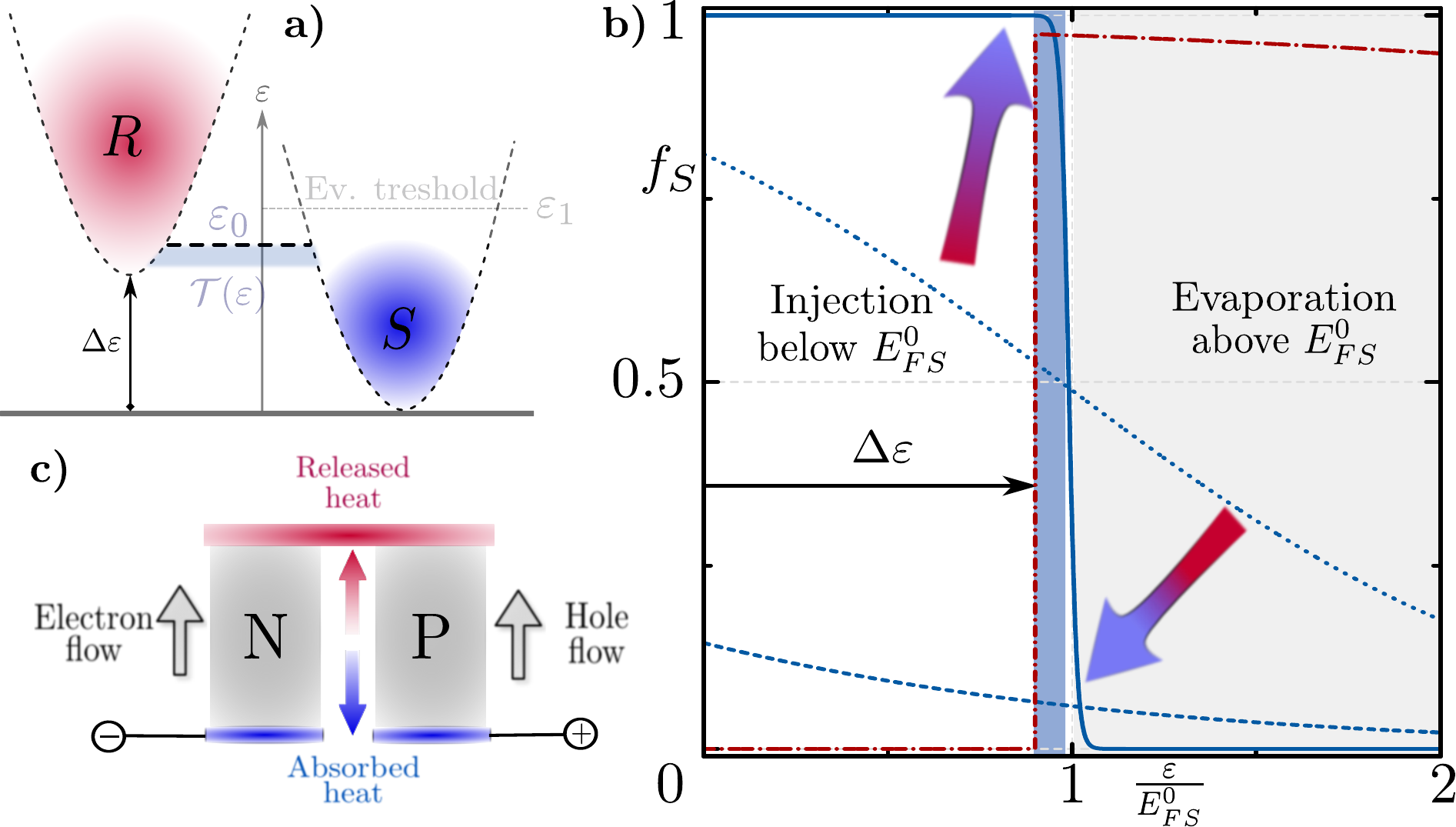}
\caption{(Color online): 
{\bf a)} Sketch of the proposed Peltier cooling scheme: 
atoms are injected from deep energy levels of the reservoir cloud ($R$) into the system cloud ($S$) just below the Fermi level through a channel 
with an energy-dependent transmission $\mathcal{T}(\varepsilon)$.  
Additionally, the system is submitted to evaporative cooling with a fixed evaporation threshold $\varepsilon_1$ above 
Fermi level, removing hot particles.  
{\bf b)} Evolution of the Fermi distribution of the system at three stages during the cooling process: initial (dashed blue curve, $T_S \approx T_{FS}$), intermediate (blue dotted curve, $T_S = 0.3 T_{FS}$) and final (solid blue curve, $T_S=0.02T_{FS}$).
The evolution, indicated by arrows, is calculated (see text) for 
$\varepsilon_1 = 1.05 E^0_{FS}$, $\gamma_{ev}\tau_0 = 1/16$, $\varepsilon_0 = 0.99E^0_{FS}$, $\Delta \varepsilon = 0.96 E^0_{FS}$ and $E^0_{FS} = 0.25 E^0_{FR}$. 
The blue and grey shaded regions indicate the injection and evaporation energy  windows, respectively. The dashed-dotted curve is the final distribution of the reservoir. 
{\bf c)} Sketch of a Peltier cooling module. 
The $n$-like and $p$-like thermoelectric materials ensure the transport of low energy (electrons) and high energy (holes) particles, 
which thus carry heat from the cold (blue) to the hot (red) region. 
}
 \label{fig:Setup}
\end{figure}

The Peltier effect is a reversible thermoelectric phenomenon, in which 
heat is absorbed or produced at the junction of two materials forming a circuit in which a current is circulated~\cite{Goldsmid}. 
%
Peltier cooling modules based on this effect are used for a variety of applications ranging from 
wine coolers to the cooling of electronic devices. 
The basic principle of a Peltier module is schematized on Fig.~\ref{fig:Setup}(c). 
It consists of two materials with Seebeck coefficients of opposite signs (p-type and 
n-type) arranged as indicated. 
A heat (entropy) current flows in both materials from the cold side to the hot side. 
Microscopically, this heat current corresponds to a flow of energetic electrons 
in the n-branch: 
hot electrons above Fermi level are `evaporated' out of the cold plate. 
Similarly, a hole current flows in the same direction in the p-branch, so that 
holes below Fermi level are being filled. Both processes lead to a rectification of the energy 
distribution of electrons in the cold plate (Fig.~\ref{fig:Setup}(b)),  
and hence to a decrease of its entropy. 

In the context of mesoscopic electronic systems, thermoelectric effects as well as thermal properties and refrigeration have recently been the 
focus of renewed interest \cite{PhysRevB.82.115314,PhysRevB.83.085428,RevModPhys.78.217}. 
The cooling of low-dimensional nanostructures has been proposed~\cite{EdwardsQDrefrigerator,PhysRevB.87.075312,PhysRevLett.112.130601} and 
experimentally realized~\cite{PhysRevLett.102.146602,PhysRevLett.102.200801,RevModPhys.78.217}, 
for example by engineering a proper energy dependence of the transmission coefficients using quantum dots. 

In the field of cold atomic Fermi gases, reaching lower temperatures is currently one of the most urgent challenges. 
Typically, the cooling of these gases is achieved using laser cooling followed by evaporative cooling~\cite{Ket1,PethickSmith}. 
Quantum degeneracy and very low absolute temperatures of the order a few hundred nano-Kelvin are typically reached with these techniques, leading to the observation of many remarkable phenomena such as the BCS-BEC crossover \cite{ZwierleinKetterleReviewFermi}. 
However, the entropy per particle, which is the relevant quantity in these well isolated fermionic gases, is still 
too large ($T/T_F\approx 0.1$)~\cite{0034-4885-74-5-054401} 
to investigate many of the most intriguing quantum effects such as the quantum Hall effect~\cite{CooperReview}, 
low-temperature transport, spin liquids~\cite{lewenstein2012ultracold}, or even antiferromagnetic order in the Hubbard model~\cite{,PhysRevLett.104.180401,RevModPhys.80.885}.

Here, we propose an efficient cooling scheme for atomic Fermi gases which uses the Peltier effect in synergy with 
evaporative cooling.
Our proposed setup is based on thermoelectric effects~\cite{Brantut08112013,2012arXiv1209.3942G,2013arXiv1306.4018H,2013arXiv1311.0769R,PhysRevLett.109.084501,2014arXiv1405.6026P,Sidorenkov:2013aa,Kim:2012aa,PhysRevA.85.063613,Karpiuk:2012aa}
and is displayed on Fig.~\ref{fig:Setup}(a). 
Two clouds of fermions, a reservoir $R$ and a system $S$ to be cooled, are prepared in harmonic traps. 
The initial Fermi energies of the two gases are $E^0_{F,R/S}=h\bar{\nu}(3N_{R/S})^{1/3}$, with $\bar{\nu}$ the average trapping frequency~\footnote{The trap frequencies 
are chosen to be identical for simplicity even though their shape could be used for further optimization of the scheme.} and 
$N_R$, $N_S$ the atom numbers ($N_R>N_S$). The lowest energy level of the reservoir is offset by $\Delta\varepsilon\geq 0$ as compared 
to that of the system.
Two processes are implemented in order to lower the entropy of the system. 
The first one is evaporative cooling at a rate $\Gamma_{ev}$ applied to particles with energy higher than a threshold $\varepsilon_1$, chosen above the Fermi level $E^0_{FS}$. 
The second simultaneous process is the injection of fermions from the reservoir into the system below the system's Fermi level which can be viewed as an `evaporation' of holes.

This is achieved by connecting the traps by a constriction~\cite{Brantut} characterized by an energy-dependent transmission $\mathcal{T}(\varepsilon)$. 
In an ideal setup, the transmission is chosen to have a box-like dependence on energy: any state 
with energy above $\Delta\varepsilon$ and below a threshold $\varepsilon_0$ located just below the Fermi level of the system 
is perfectly transmitted (Fig.~\ref{fig:Setup}(a)).  
%

The combination of these two processes, the evaporative cooling and the injection of fermions below the Fermi surface, induces an efficient cooling. 
This can be seen from the time evolution of the energy distribution displayed on Fig.~\ref{fig:Setup}(b). 
Starting initially from a broad hot distribution, it evolves towards a rectified distribution with a sharp drop at the 
Fermi level, characteristic of a low temperature. The parameters of the cooling process can be chosen, such that the atom number in the system changes 
only slightly, since the atom losses from evaporative cooling can be compensated by the injection of the reservoir atoms. 
At the same time, the reservoir is heated and looses atoms. 
However, the bottom of the energy distribution of the reservoir remains filled (Fig.~\ref{fig:Setup}(b)), ensuring an efficient injection of cool particles. 
The cooling process stops when the system energy distribution becomes equal to the reservoir distribution in the transmission window. 
We show below that both the final entropy per atom and the cooling rate are improved in comparison to evaporative cooling only, 
by approximately a factor of four.  

\paragraph{Model of the cooling process} We describe the process in terms of coupled rate equations for the distribution functions $f_S$ and $f_R$.
We assume that thermalization in the system and in the reservoir is fast, so that they can be considered to be in thermodynamic equilibrium. 
Under this assumption, the particle current leaving the reservoir is given by the Landauer formula:  
\begin{eqnarray}
 I_N & = & \frac{1}{h} \int \textrm{d}\varepsilon\; \mathcal{T}(\varepsilon) \left[ f_R(\varepsilon)-f_S(\varepsilon)\right] \nonumber \\
 {} &=& - \int \textrm{d}\varepsilon g_R(\varepsilon)\frac{df_R}{dt}(\varepsilon) = \int \textrm{d}\varepsilon g_S(\varepsilon)\frac{df_S}{dt}(\varepsilon)\,.\nonumber
\end{eqnarray}
In this expression,  $g_R(\varepsilon) = \frac{(\varepsilon-\Delta \varepsilon)^2}{3(h\bar{\nu})^3}\vartheta(\epsilon-\Delta 
\varepsilon)$ and $g_S(\varepsilon) = \frac{\varepsilon^2}{{3(h\bar{\nu})^3}}\vartheta(\varepsilon)$ are the density of states 
in the reservoir and in the system, with $\vartheta$ the Heaviside function. 
The coupled evolution of the two distribution functions is thus given by~: 
\begin{eqnarray}
 g_R(\varepsilon)\frac{df_R(\varepsilon)}{dt} & = & -\frac{\mathcal{T}(\varepsilon)}{h}\left[f_R-f_S\right](\varepsilon)\label{eq:Evolution_transport1}\\
 g_S(\varepsilon)\frac{df_S(\varepsilon)}{dt} & = & \frac{\mathcal{T}(\varepsilon)}{h}\left[ f_R-f_S\right](\varepsilon) - \Gamma_{ev}(\varepsilon)g_S(\varepsilon)f_S(\varepsilon)\,,
 \label{eq:Evolution_transport2}
\end{eqnarray}
In the second equation, the effect of evaporation has been included as a leak of high energy particles above a fixed energy threshold $\varepsilon_1$, with an 
energy independent rate $\Gamma_{ev}(\varepsilon) = \gamma_{ev}\vartheta(\varepsilon-\varepsilon_1)$~\cite{Ket1}. 
Since $\mathcal{T}(\varepsilon)$ is dimensionless, the typical time-scale that rules the time-evolution in these equations is  
$\tau_0 = hg_S(E^0_{FS})=h\left(E^0_{FS}\right)^2/3(h\bar{\nu})^3$. This is the compressibility divided by the `conductance' quantum, which 
has been identified as the time-scale for the particle transport in previous experiments~\cite{Brantut,Brantut08112013}.

In principle the scattering of the atoms has to be taken into account in the evolution equations. 
However, since we assume the scattering to be the fastest time-scale in the problem, we take it into account as an instantaneous re-thermalization. 
The evolution of the system is implemented by time-evolving the equations~\eqref{eq:Evolution_transport1} and \eqref{eq:Evolution_transport2} for a small time step $\delta t$. From the particle numbers $N_{S,R}(t+\delta t)$ and 
energies $E_{S,R}(t+\delta t)$ one obtains new values of the chemical potentials $\mu_{S,R}$ and temperatures $T_{S,R}$ assuming thermodynamic equilibrium for a non-interacting gas.  
The resulting equilibrium Fermi functions $f_S(t+\delta t,\varepsilon)$ and $f_R(t+\delta t,\varepsilon)$ are subsequently evolved using the rate equations and the entire procedure is repeated. This description of cooling is in line with the pulsed approach of evaporation developed for example in Ref.~\cite{DavisMewesKetterle,Ket1}. 
The time-step $\delta t$ is chosen small enough, such that it does not affect the resulting evolution. 

Unless specified, a transmission of the form $\mathcal{T}(\varepsilon) = \vartheta(\epsilon_0-\varepsilon)$ will be considered.  
This means that only states in the energy window 
$\varepsilon\in[\Delta \varepsilon,\varepsilon_0]$ 
can be transmitted (`box-like' transmission). 
As recently pointed out for electronic mesoscopic systems~\cite{PhysRevLett.112.130601} and 
further discussed in the last section, this transmission is the optimal choice to achieve the best cooling performances.
To summarize, the scheme involves three characteristic energy thresholds: 
$\Delta\varepsilon\leq\varepsilon_0\leq\varepsilon_1$, corresponding respectively to the 
energy offset between $R$ and $S$, the maximum injection energy into $S$, 
and the minimum evaporation energy out of $S$.  

\paragraph{Results and comparison to evaporative cooling}

\begin{figure}[ht!]
 \includegraphics[width=0.9\linewidth]{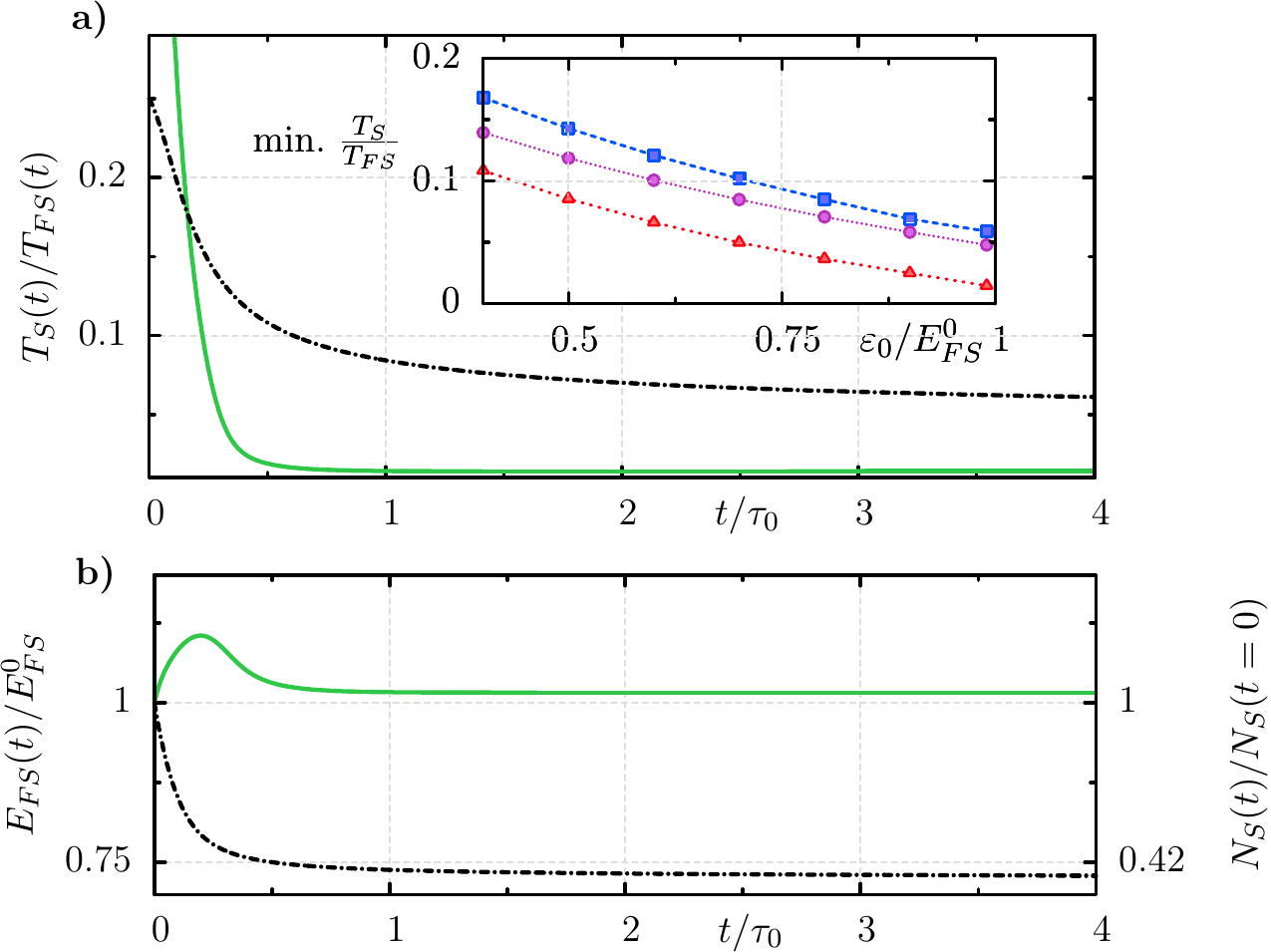}
 \caption{(Color online): Evolution during the cooling process of 
  {\bf a)} the entropy per particle $T_S(t)/T_{FS}(t)$ 
  and {\bf b)} the Fermi energy $E_{FS}(t)$ (left axis) and particle number $N_S(t)$ (right axis). 
  The solid curves are for the Peltier cooling with $E^0_{FS}/E^0_{FR}=1/4$ and $\varepsilon_1 = 1.05E^0_{FS}$, $\gamma_{ev}\tau_0 = 1/16$, 
  $\varepsilon_0 = 0.99E^0_{FS}$, $\Delta \varepsilon = 0.96E^0_{FS}$. 
  The dashed-dotted curves are for evaporative cooling only, with an initial particle number $N=N_S+N_R$ and $\varepsilon_1 = 1.05E_{F}$. 
  Inset: minimum entropy per particle achieved by Peltier cooling, as a function of $\varepsilon_0$, for : 
 $\varepsilon_1 = 1.05 E^0_{FS}$ and $(\epsilon_0-\Delta\varepsilon)/E^0_{FS} = 3\%$ (red triangles), $20\%$ (purple circles) and  
 $\varepsilon_1 = 1.5 E^0_{FS}$, $(\epsilon_0-\Delta\varepsilon)/E^0_{FS} = 3\%$ (blue squares).
 }
 \label{fig:Temperature_particle_number}
\end{figure}

In the following we demonstrate the potential of the proposed Peltier cooling scheme by comparing it to commonly used evaporative cooling. 
We focus on the reachable entropy per particle 
 and on the cooling rate.
The entropy per particle $s$ in a trapped non-interacting Fermi-gas is related to the ratio $T/T_F$ by $s = \frac{\pi^2k_B}{2} \frac{T}{T_F}$ at low temperature, 
so that we use equivalently $s$ or $T/T_F$ below. 
The evolution of the entropy per particle 
is shown in the upper panel of Fig.~\ref{fig:Temperature_particle_number}. 
Assuming that the two gases are first prepared from a single cloud by evaporative cooling, 
we used a typically reached entropy per particle of $T/T_F=0.25$ as an initial value (with $T_F$ the Fermi temperature of the total cloud). 
Initially, the ratio of the atom numbers between the system and the reservoir is chosen such that   
$E^0_{FS}/E^0_{FR}=N_S/N_R = 0.25$, which leads to the initial entropy per atom of $T_S/T_{FS} \simeq 1.07$ and $T_R/T_{FR} \simeq 0.27$. For the results in Fig.~\ref{fig:Temperature_particle_number} we have chosen the evaporation threshold $\varepsilon_1 = 1.05E^0_{FS}$, 
the maximal transmission energy $\varepsilon_0 = 0.99E^0_{FS}$, i.e.~close to the target Fermi energy (which is the initial one) and the chemical potential 
bias $\Delta \varepsilon = 0.96E^0_{FS}$. The evaporation rate is chosen to be $\gamma_{ev}\tau_0 = 1/16$. 
Fig.~\ref{fig:Temperature_particle_number} shows that a very efficient reduction of the entropy per particle to a 
value of approximately $T_S/T_{FS}\approx 0.02$ is achieved within a short time-scale. At longer times, a slight rise in the entropy per particle sets in. The number of particles in the system is approximately constant with 
a slight increase at short times. This increase in the atom number in the system shows that during this time, the injection of fermions from the reservoir is particularly efficient. This increase is reduced by a dominating evaporation at intermediate times, until a quite stable state is reached at longer times.

\begin{figure}
 \includegraphics[width=0.7\linewidth]{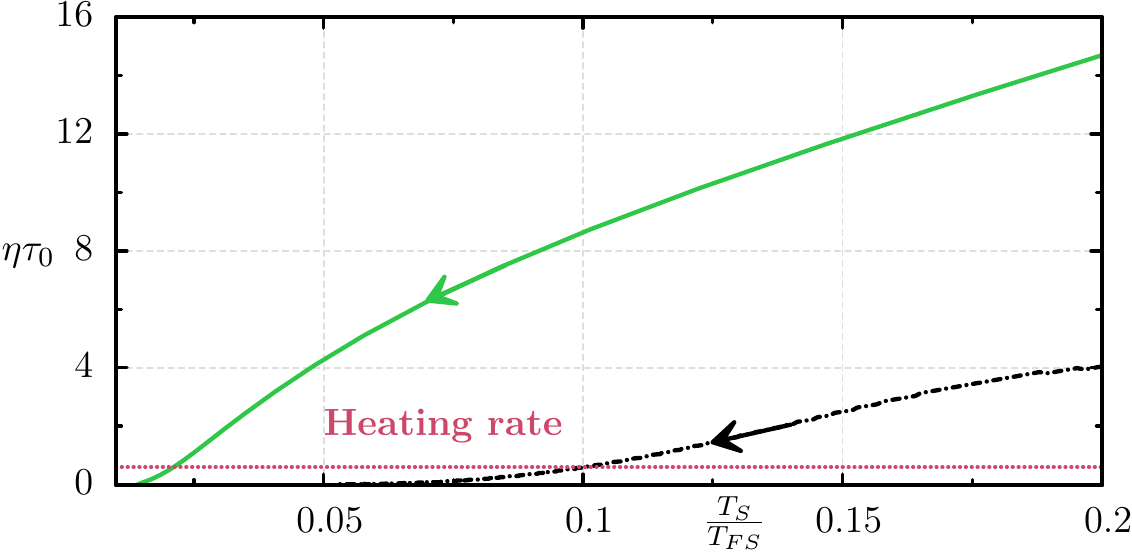}
 \caption{(Color online): Dimensionless cooling rate $\eta(t)\tau_0$ as a function of $T_S/T_{FS}$, for the same parameters as in Fig.~\ref{fig:Temperature_particle_number}. 
 The black dashed curve is for evaporative cooling only. Arrows indicate the direction of the time evolution. The horizontal (red) dotted line indicates a typical heating rate limiting these cooling processes.}
 \label{fig:Rate}
\end{figure}

In experimental setups, various heating mechanisms such as spontaneous emission~\cite{Footbook},  
as well as particle losses, can limit drastically the lowest entropy that can be reached. 
The cooling process will stop being efficient when the cooling rate becomes comparable to the heating/emission rate, 
and it is therefore important to compare these two rates.  
To this purpose, we define the cooling rate $\eta$ as the time derivative of the entropy per particle~:
$ \eta (t) = -\frac{d s_S(t)}{dt}\,,$
and display it in Fig.~\ref{fig:Rate} versus the corresponding value of $T_S(t)/T_{FS}(t)$. 
The horizontal dashed line stands for a typical value of the experimental heating rate. From this plot, one can directly 
read off the entropy per particle which can be reached (here about $0.03T_S/T_{FS}$) in the presence of this heating rate. 

To assess the usefulness of our cooling scheme, we compare it to evaporative cooling applied to the total initial cloud with $N=N_R+N_S$ 
with the same initial temperature, here  $T/T_F = 0.25$, and the same evaporation threshold relative to the Fermi 
energy $\varepsilon_1 = 1.05E^0_{F}$, where $E^0_F$ is the initial Fermi energy of the total cloud. Since for the evaporative cooling no separation of the total cloud in two subclouds is performed we use the index $S$ to label the quantities of the entire cloud. 
We see from Fig.~\ref{fig:Temperature_particle_number} that at short times, the entropy reached by the proposed cooling scheme is much lower than the one reached by evaporative cooling. So if one aims at reaching a given low value of $T/T_F$, this value is reached faster with the proposed scheme.  
At the same time one sees that the particle number during the evaporative cooling is reducing drastically to about 40 percent of its 
initial value~\footnote{However, note that this value is still larger than the particle number in the system gas.}.
At infinite times, the evaporative cooling would empty the reservoir and during this process reach lower and lower entropy per particles. 
 Nevertheless, as cooling takes place, evaporation is less and less efficient, and
the cooling rate of evaporation slows down considerably:  
as seen on Fig.~\ref{fig:Rate} the cooling rate for the Peltier scheme is much larger than for the evaporative scheme,  
at a given entropy per particle. 
%
Due to this faster cooling rate, the temperatures which can be reached in the presence of heating or spontaneous emission 
are much deeper in the degenerate regime when using the Peltier cooling. 
As seen on Fig.~\ref{fig:Rate}, for the chosen heating rate, an entropy per particle of order $0.03 T_S/T_{FS}$ is reached 
using the Peltier scheme, in contrast to $0.13 T_S/T_{FS}$ using evaporative cooling only. 

The inset of Fig.~\ref{fig:Temperature_particle_number}(a) illustrates how the minimum of the entropy per particle depends on the parameters of the setup. The injection energy window $\varepsilon_0-\Delta \varepsilon$ should be relatively narrow to obtain a low entropy, but broad enough 
to allow for a fast cooling. Having in addition an evaporation threshold as close as possible to the target Fermi
energy also improves the final value of the entropy per particle. 
The cooling scheme could further be optimized by changing some of the parameters in a time-dependent manner, 
as commonly done for the threshold in evaporative cooling. 
For the sake of simplicity, we kept all parameters constant in our study.

\begin{figure}
\includegraphics[width=0.9\linewidth]{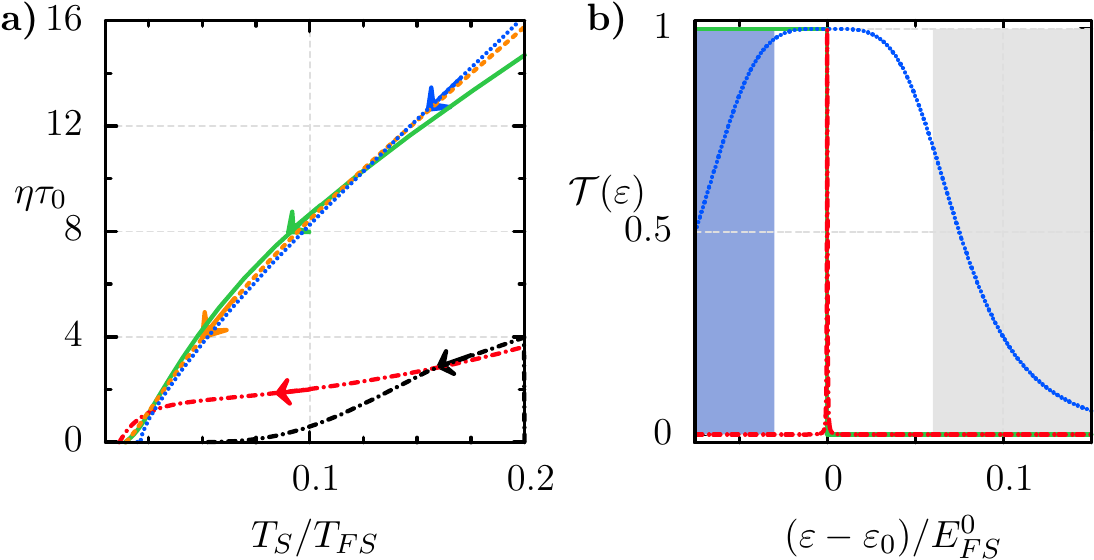}
\caption{(Color online) {\bf a)} : Dimensionless cooling rate $\eta(t)\tau_0$ as a function of $T_S/T_{FS}$, for  $\Delta\varepsilon = 0.96E^0_{FS}$ and various transmissions centered at $\varepsilon_0 = 0.99 E^0_{FS}$ (the other parameters are taken as in Fig.~\ref{fig:Temperature_particle_number}): 
The (red) dot-dashed and (orange) dashed curve correspond to a single and 100 parallel resonant level(s), respectively, with $\Gamma = 2\cdot 10^{-4} E^0_{FS}$. The (blue) dotted curve is for two resonant levels in series of width $\Gamma = 0.3 E^0_{FS}$ and the (green) solid curve is for an ideal box transmission. (Black) dashed-dotted curve shows the evaporative cooling only. 
{\bf b)}: The corresponding energy-dependent transmission coefficients. The grey area indicates states above $\varepsilon_1$, while the blue one indicates those below $\Delta \varepsilon$, which do not participate to transport.
}
\label{fig:T_vs_eta}
\end{figure}

\paragraph{Possible implementations of the energy-dependent transmission}
The Peltier cooling scheme relies on a transmission coefficient ensuring proper `energy filtering' between the two gases. We now discuss possible realizations of appropriate transmission functions  with state-of-the-art  
projection techniques~\cite{1367-2630-13-4-043007,GreinerSAI,BlochSAI}. We consider mainly two distinct forms (see Fig.~\ref{fig:T_vs_eta}(b)). First, a narrow (delta-function like) transmission in energy which can for example be realized by a single resonant level (or many in parallel)~\cite{datta}. Such a narrow energy filter~\cite{Mahan23071996}, has been predicted to achieve the maximization of the cooling efficiency or equivalently of the thermoelectric figure of merit. 
Second, an approximately box-like transmission realized by two such resonant levels in series as discussed in the mesoscopics 
context \cite{PhysRevLett.112.130601,Flindtarxiv}. Such a box-like transmission with a finite width in energy is expected to show the maximum cooling power and best cooling rate (see Ref.~\cite{PhysRevLett.112.130601} and supplementary material~\cite{Supp}). 
 
The results for the cooling rates as a function of $T_S/T_{FS}$  are displayed in Fig.~\ref{fig:T_vs_eta}(a). As predicted, the idealized box-like transmission 
(solid green line) allows for reaching a very low temperature with a fast cooling rate. Two quantum dots connected in series (blue dotted curve in Fig.~\ref{fig:T_vs_eta}(a)) 
realize a good approximation to the box-like transmission and achieve a final entropy per particle which is only slightly higher. 
We also considered a single resonant level (red dash-dotted curve), with a width such that a low final entropy comparable to that 
of the idealized box is achieved. In contrast to the latter, this leads to a much slower cooling rate, comparable to that of evaporative 
cooling~\footnote{The lowest entropy per particle which can be reached for the single resonant level is mainly determined by its width $\Gamma$ and can be decreased 
considerably by lowering its width on the cost of slowing down its rate.}. 
Only in the final cooling stage, when the entropy per particle is low, does the cooling rate of the single resonant level become comparable to that of the ideal box. 
This low cooling rate can to some extent be overcome by using many resonant levels in parallel (orange dashed curve): 
a fast initial cooling is then observed, comparable to that of the idealized box, with only a slight increase of the minimal entropy per particle which can be reached. 
In summary, we have identified two different ways of realizing efficient cooling, either by connecting relatively broad resonant levels in series, or 
by using a large number of narrow resonant levels in parallel. 
The attainable values of $T_S/T_{FS}$ remain in all considered cases significantly better than what can be achieved from evaporation only.\\

\paragraph{Conclusion}
In this article, we have introduced a Peltier cooling scheme for fermionic gases, which combines conventional evaporation 
with energy-selective injection of particles. In a nutshell, this scheme can be described as a simultaneous evaporative 
cooling of particles and holes. We have proposed different realizations of the proper energy filtering between the 
reservoir and the cooled system, in line with the recent development of mesoscopic-like channels in cold atom gases~\cite{Brantut,ThywiessenQPC}. 
The proposed scheme achieves fast and efficient cooling down to temperatures deep in the 
quantum degenerate regime, a much desired current goal in the field of atomic fermion gases. 
The present work also demonstrates that the recent fundamental studies of coupled particle 
and entropy transport in cold atomic gases~\cite{Brantut08112013,2012arXiv1209.3942G,2013arXiv1306.4018H,2013arXiv1311.0769R,PhysRevLett.109.084501,2014arXiv1405.6026P,Sidorenkov:2013aa,Kim:2012aa,PhysRevA.85.063613,Karpiuk:2012aa} may also have useful implications for further experimental developments of the field. 


\acknowledgments
We thank J.-P.~Brantut,  M. B\"uttiker, T.~Esslinger, S.~Krinner, H. Moritz, D.~Papoular, J.~L.~Pichard, B.~Sothmann, S.~Stringari and R.~S.~Whitney 
for useful discussions and suggestions. 
Support was provided by the DFG, the DARPA-OLE program, NCCR QSIT, and the FP7 project Thermiq. 
\bibliography{Paper_cooling_biblio} 

\newpage

\begin{center}
\bf Supplementary material for : 'Peltier Cooling of Fermionic Quantum Gases'
\end{center}

\section{Derivation of the main equations}

Equations (1) and (2) in the main text rule the evolution of the distribution functions $f_R$ and $f_S$ in the reservoir and system respectively, under the influence
of an energy dependent coupling represented by a transmission probability $\mathcal{T}(\varepsilon)$ and a 'leak' of high energy particles 
representing the effect of evaporation.\\
The starting point to derive these equations is the balance of particle currents~:
\begin{eqnarray}
 \dot{N_R} &=& I_{S \rightarrow R} - I_{R \rightarrow S}\\
 \dot{N_S} &=& I_{R \rightarrow S} - I_{S \rightarrow R}-I_{evaporation}\, 
\end{eqnarray}
which expresses the variation of the particle numbers in the system $S$ and reservoir $R$. Note that the evaporation which acts on the system leads to particle losses, such that the total particle number is not conserved. The Landauer formula gives the expression for the current flow
between $S$ and $R$~:
\begin{equation}
 I_{tot} = I_{S \rightarrow R} - I_{R \rightarrow S} = \frac{1}{h}\int d\varepsilon \,\mathcal{T}(\varepsilon)\,\left[f_S(\varepsilon,t) - f_R(\varepsilon,t)\right]\,.
\end{equation}
The distributions fulfill the following relations~:
\begin{eqnarray}
 \int d\varepsilon \,g_{S,R}(\varepsilon)\,f_{S,R} (\varepsilon,t) &=& N_{S,R}(t)\\
 \int d\varepsilon \,\varepsilon\, g_{S,R}(\varepsilon)\,f_{S,R} (\varepsilon,t) &=& E_{S,R}(t)\,
\end{eqnarray}
where $N_{S,R}$ and $E_{S,R}$ are the particle number and energy in $S,R$ at time $t$.\\

The evaporation term has the following expression~:
\begin{equation}
 I_{evaporation} = \int d\varepsilon \, g_{S}(\varepsilon)\,f_{S}(\varepsilon)\,\Gamma_{ev}(\varepsilon)\,,
\end{equation}
with $\Gamma_{ev}(\varepsilon) = \gamma_{ev} \theta(\varepsilon-\varepsilon_1)$, representing the evaporation above a fixed threshold $\varepsilon_1$
at a rate $\gamma_{ev}$.

Finally, expressing the variation of particle number in $S$ and $R$ as
\begin{equation}
 \dot{N}_{S,R} = \int d\varepsilon \,g_{S,R}(\varepsilon)\,\frac{d f_{S,R}(t)}{dt}(\varepsilon)
\end{equation}
gives the expressions (1) and (2) of the main text.

\section{Maximization of power factor}

In the linear response regime, one can provide a quantitative explanation for the optimization of the cooling efficiency provided by a box-like transmission.
The arguments that we will develop are similar to those presented in Ref.~~\cite{PhysRevLett.112.130601} in the mesoscopic context which covers also the nonlinear regime.
\\

Here, we assume the response to be linear and search for the energy dependent transmission probability $\mathcal{T}(\varepsilon)$ that optimizes 
the power factor $\mathcal{P} = \alpha^2G$~\cite{macdonaldthermoelectricity}, where $\alpha$ is the Seebeck coefficient and $G$ the conductance, given by~:
\begin{eqnarray}
 G & = & \frac{1}{h}\int_0^\infty d\varepsilon \, \mathcal{T}(\varepsilon)\left(-\frac{\partial f}{\partial \varepsilon} \right)\\
 \alpha & = & \frac{1}{T}\frac{\int_0^\infty d\varepsilon\, \mathcal{T}(\varepsilon)(\varepsilon-\mu)\left(-\frac{\partial f}{\partial \varepsilon} \right)}{\int_0^\infty d\varepsilon \, \mathcal{T}(\varepsilon)\left(-\frac{\partial f}{\partial \varepsilon} \right)}\,.
\end{eqnarray}

The previous expressions give the following result for the power factor~:
\begin{equation}
 \mathcal{P} = \frac{k_B^2}{h} \cdot \frac{\left[\int du \mathcal{T}(u)\frac{u-\xi}{4\cosh^2{\left((u-\xi)/2\right)}} \right]^2}{\int du \frac{\mathcal{T}(u)}{4\cosh^2{\left((u-\xi)/2\right)}}}\,,
\end{equation}
with $\xi=\beta\mu$. Then, the previous expression is optimized with respect to the function $\mathcal{T}(u)$, under the constraint that for all values of $u$ the transmission fulfills $0 \leq\mathcal{T}(u)\leq 1$.
A numerical solution to this functional problem results in a box-like function for the transmission $\mathcal{T}(u)$, with a threshold close to the dimensionless 
chemical potential $\xi = \beta\mu$.\\

One can get further insight using the Mott-Cutler formula, which provides a low temperature expression for the Seebeck coefficient : $\alpha = \frac{k_B\pi^2}{3}\frac{\mathcal{T}'(\mu)}{\mathcal{T}(\mu)}$.
Then, the power factor becomes 
\begin{equation}
 \mathcal{P} = \frac{k_B^2\pi^4}{9} \frac{(\mathcal{T}'(\mu))^2}{\mathcal{T}(\mu)}\,.
\end{equation}
This expression points out, that not only the actual value of the transmission close to the chemical potential is important, but also its derivative. The box-like transmission obtained by the numerical optimization has a large derivative close to the chemical potential. 

As also noticed in~\cite{PhysRevLett.112.130601}, the optimization of the power factor leads to a different transmission than the one obtained under the 
condition of optimized efficiency, which would result in a sharp energy resolved transmission~\cite{Mahan23071996}.

\section{Transmission for two resonant levels in series}

The transmission probability as a function of energy for two resonant levels in series connected to two identical reservoirs $L$ and $R$
is given by the Fisher-Lee formula\cite{datta,PhysRevB.23.6851}, which relates the transmission $\mathcal{T}$ to the Green's function $\mathcal{G}$~:
\begin{equation}
 \mathcal{T} (\varepsilon) = Tr\left( \Gamma_L\cdot \mathcal{G} \cdot \Gamma_R \cdot \mathcal{G}^\dagger\right)\,,
\end{equation}
where $\Gamma_L$ and $\Gamma_R$ contain the tunneling rates from the reservoirs to the resonant levels : $\Gamma_L = \left( \begin{array}{cc} \Gamma & 0 \\ 0 & 0\end{array}\right)$ 
and $\Gamma_R = \left( \begin{array}{cc} 0 & 0 \\ 0 & \Gamma\end{array}\right)$. In the previous expression for the transmission, the resonant levels' Green's function
is given by $\mathcal{G} = \left[\varepsilon\mathbb{I}-H-\Sigma\right]^{-1}$, with~:
\begin{equation}
 H = \left(
 \begin{array}{cc}
  \varepsilon_0 & t \\
  t & \varepsilon_0
 \end{array}
 \right)
\end{equation}
the Hamiltonian of the two identical resonant levels at energy $\varepsilon_0$ connected by a tunnel barrier of amplitude $t$, and $\Sigma = -i\frac{\Gamma_L+\Gamma_R}{2}$ the imaginary part of the self-energy.
Then, the transmission reads~:
\begin{equation}
 \mathcal{T}(\varepsilon) = \frac{\Gamma^2t^2}{\left[(\varepsilon-\varepsilon_0)^2+\Gamma^2/4-t^2\right]^2+\Gamma^2t^2}.
\end{equation}
One sees that choosing $t = \Gamma/2$ ensures a maximal transmission of $1$ for $\varepsilon = \varepsilon_0$, and that the effective width of the transmission
resonance is given by $\Gamma/\sqrt{2}$. Furthermore, the obtained transmission has a sharper energy dependence than the one for a single dot, and then is closer to the 
ideal (box-like) transmission function, as also noticed in~\cite{PhysRevLett.112.130601}.

\end{document}